\documentclass[twocolumn,amsfonts,superscriptaddress]{revtex4-1}

\usepackage{amsmath}

\pdfoutput=1

\begin{document}

% ----------------- text below here -------------

\title{Vacuum energy is non-positive for $(2+1)$-dimensional holographic CFTs}

\author{Andrew Hickling}
\affiliation{Theoretical Physics Group, Blackett Laboratory, Imperial College, London SW7 2AZ, UK } 
\author{Toby Wiseman}
\affiliation{Theoretical Physics Group, Blackett Laboratory, Imperial College, London SW7 2AZ, UK }

\date{August 2015}

\begin{abstract}
We consider a $(2+1)$-dimensional holographic CFT on a static spacetime with globally timelike Killing vector. Taking the spatial geometry to be closed but otherwise general we expect a non-trivial vacuum energy at zero temperature due to the Casimir effect. 
We assume a thermal state has an AdS/CFT dual description as a static smooth solution to gravity with a negative cosmological constant, which ends only on the conformal boundary or horizons. 
A bulk geometric argument then provides an upper bound on the ratio of CFT free energy to temperature.
Considering the zero temperature limit of this bound implies the vacuum energy of the CFT is non-positive. 
Furthermore the vacuum energy must be negative unless  the boundary metric is locally conformal to a product of time with a constant curvature space.
We emphasise the argument does not require the zero temperature bulk geometry to be smooth, but only that singularities are `good' so are hidden by horizons at finite temperature.

\end{abstract}

\maketitle

%-----------------------------------------------------------------
%
\section{Introduction}
%
%-----------------------------------------------------------------

The AdS/CFT correspondence \cite{Maldacena:1997re} provides a very elegant reformation of certain questions concerning specific strongly coupled CFTs in terms of classical geometric calculations in a dual gravity theory. Such methods are particularly powerful for questions related to CFTs deformed by sources in a manner that depends on spacetime where first principle computations in such CFTs are very challenging. An example of this is considering the CFT on curved spacetime as reviewed in \cite{Marolf:2013ioa}.

Here we examine a basic property of quantum field theories on curved static spacetimes with globally timelike Killing vector, namely their vacuum energy. 
This was discussed in \cite{Horowitz:1998ha,Myers:1999psa} in the context of AdS/CFT where it was pointed out that the bulk stress tensor determined by holographic renormalisation \cite{Henningson:1998gx,Balasubramanian:1999re,deHaro:2000xn,Skenderis:2002wp}, which as usual is chosen to vanish on flat space, then gives the vacuum or Casimir energy when the boundary space is non-trivial
and the bulk geometry is that corresponding to the vacuum state.
In this work we consider $(2+1)$-dimensional holographic CFTs on such spacetimes which due to the absence of a conformal anomaly may be chosen without loss of generality to have ultrastatic form. We  take the spatial geometry, $\Sigma$, to be closed so that the total vacuum energy is finite, and being in odd dimensions is scheme independent in the holographic renormalisation.
This energy will be a functional of the 2-dimensional spatial geometry $\Sigma$ which, other than being closed, may have general topology and metric. The purpose of this paper is to show that under reasonable assumptions on the nature of the bulk geometry dual to the vacuum state, simple geometric considerations 
lead to the conclusion that the CFT vacuum energy is non-positive for any closed space $\Sigma$, and negative unless $\Sigma$ has constant curvature (so is locally a sphere, torus or hyperbolic space).

To put this result in context one can consider whether such a result applies to the vacuum energy in other theories. 
For $(1+1)$-dimensional CFTs a closed spatial geometry $\Sigma$ is simply a circle and so there is no interesting local geometry. Then a classic computation yields that the Casimir energy is a function of the circle size and the central charge. Likewise the Casimir energy may be simply computed for free theories too.
However for more spatial dimensions, such as the $(2+1)$-dimensional case we consider here, then $\Sigma$ may have complicated local geometry that the vacuum energy depends on. One might imagine that free field theories would allow computations to be performed. For example one could consider a  free scalar field (massless, massive or conformally coupled), on such an ultrastatic spacetime. The vacuum energy is then related to the functional determinant of an elliptic operator on the space $\Sigma$ \cite{Birrell:1982ix,Fulling:1989nb}. However such determinants are very subtle as they are naively divergent and must be regulated. Then even computing the vacuum energy for highly symmetric spaces is a very non-trivial  task, albeit a well defined one in odd dimensions where there is no scheme dependence \cite{Candelas:1983ae}.
With the simplest cases being so challenging it is not surprising that, to our knowledge, results such as bounds on the vacuum energy for general $\Sigma$ do not exist even for free matter above $(1+1)$-dimensions. Thus holography provides a very powerful tool for calculation, allowing elementary methods to give global results on the vacuum energy for strongly coupled CFTs as a functional of the space $\Sigma$.

One assumption we make is that the vacuum state is described by a static bulk geometry which is smooth when heated up to any finite temperature. We do not require the vacuum dual geometry to be smooth at zero temperature, and note that in important canonical examples it is not. While there is some control mathematically over existence of infilling bulk geometries for given boundary spaces (for example \cite{GrahamLee,Anderson:2002xb}) this is not generally a solved problem, and particularly given that at zero temperature the bulk may be singular, is presumably very hard to understand generally.  Thus for us it will remain an assumption. Technically we will prove an inequality involving the free energy at finite temperature where we assume the bulk is smooth, with any zero temperature singularities being `good' and hence shrouded by horizons \cite{Gubser:2000nd}. Then we take the zero temperature limit of this  to derive that the vacuum energy is non-positive. 

In fact such a bound on free energy was obtained for holographic CFTs in the special case where $\Sigma$ has constant scalar curvature (so a round sphere, flat torus or quotients thereof, or compact hyperbolic spaces) in \cite{Galloway:2015ora} using similar methods but the mass definition of \cite{WangMass} (see also the related earlier works \cite{Boucher:1983cv,Chrusciel:2000az} in the same constant scalar curvature setting but without horzions).
Consequently this does not allow study of the Casmir energy when the boundary metric is deformed, but instead gives a general statement about the  properties of thermal states for the CFT on these specific spaces. In this work we use the holographic stress tensor to  compute the vacuum energy, allowing us to consider any boundary space $\Sigma$. In fact while the thermal inequality in  \cite{Galloway:2015ora} applies in general dimension, interestingly it is only in the case of $(2+1)$ boundary dimensions that it can be extended to general spaces $\Sigma$ and hence derive a global statement on the vacuum energy as a functional of $\Sigma$.

The structure of the paper is as follows. In section~\ref{sec:holo} we will review the geometric dual description of a holographic CFT, discuss the extraction of the boundary stress tensor, and outline the assumptions we make of the bulk geometry. Then in section~\ref{sec:bulk} we consider the bulk equations and introduce the key geometric tool we will use to control the vacuum energy, namely that the bulk optical Ricci scalar is a super-harmonic function, allowing an inequality relating boundary terms at the conformal boundary and any bulk horizons. In the following sections~\ref{sec:boundary} and~\ref{sec:horizon} we compute these boundary terms, and collecting these in~\ref{sec:result} arrive at a thermodynamic inequality involving the free energy. Taking the zero temperature limit then yields our claimed result that the zero temperature vacuum energy is non-positive. We conclude with a brief discussion in section~\ref{sec:discussion}.
\\

\noindent
\emph{Note added:} After completion of this work an alternate derivation of our result was pointed out to us by Juan Maldacena. This uses an elegant geometric bound related to the Gauss-Bonnet theorem \cite{Anderson}. For the interest of the reader we have added section~\ref{sec:Anderson} detailing this alternate approach.

%-----------------------------------------------------------------
%
\section{The dual bulk geometry
\label{sec:holo}}
%
%-----------------------------------------------------------------

We restrict our attention to the universal gravity sector of the AdS/CFT correspondence (see for example the review \cite{Marolf:2013ioa}), so that the zero temperature vacuum and thermal states of the $(2+1)$-dimensional CFT in the absence of sources are described by solutions of the $(3+1)$-dimensional pure gravity Einstein equations with negative cosmological term. Let the bulk metric be $g^{(4)}$, then it must satisfy the Einstein condition,
\begin{eqnarray}
R^{(4)}_{\mu\nu} = - \frac{3}{\ell^2} g^{(4)}_{\mu\nu}
\end{eqnarray}
where $\ell$ determines the AdS length, and is related to $c$, the CFT `effective central charge', as $c = \ell^{2} / 16 \pi G_{(4)}$, where $G_{(4)}$ is the bulk Newton constant. For the CFT to be well described by a semi-classical gravity dual we require $c \gg 1$.

The AdS/CFT correspondence dictates that the metric $g^{(4)}$ has a conformal boundary whose geometry gives the conformal class of the spacetime the CFT lives on. Let us denote the metric of the CFT's spacetime as $g_{CFT}$. The asymptotic approach to this conformal boundary then determines the expectation value of the renormalised CFT stress tensor, $T_{CFT}$, and hence the energy of the state that is described by this bulk geometry.

Consider the CFT on a general static closed spacetime,
\begin{eqnarray}
g_{CFT} = - N(x) dt^2 + \bar{g}_{ab}(x) dx^a dx^b
\end{eqnarray}
where $(\Sigma, \bar{g})$ is a smooth closed $2$-dimensional Riemannian manifold with  local coordinates $x^a$, and
$\partial / \partial t$ is a globally timelike Killing vector, so that $N$ is a positive function over $\Sigma$.
Since the Killing vector is globally timelike, and so $N > 0$, we can move to an ultrastatic conformal frame. There is no conformal anomaly in $(2+1)$-dimensions, so the stress tensor transforms simply under the required conformal transformation. We take a new frame, $g'_{CFT} = \Omega^2 g_{CFT}$, where $\Omega^2 = 1/N$, and then the metric in this frame is ultrastatic, 
\begin{eqnarray}
g'_{CFT} = - dt^2 + \bar{g}'_{ab}(x) dx^a dx^b
\end{eqnarray}
with $\bar{g}'_{ab} = \bar{g}_{ab} / N$, and the stress tensor in the new frame is,
\begin{eqnarray}
T'_{CFT} =  \Omega^{-1} T_{CFT}  = \sqrt{N} T_{CFT} \; .
\end{eqnarray}
In particular we will be interested in the CFT energy, $E$, defined with respect to the time translation Killing vector $v = \frac{\partial}{\partial t}$. 
This energy is given as,
\begin{eqnarray}
E = \int_{\Sigma} \sqrt{\bar{g}} T^{CFT}_{AB} n^A v^B = \int_{\Sigma} \sqrt{\bar{g}} \frac{1}{\sqrt{N}} T^{CFT}_{tt} 
\end{eqnarray}
where $x^A = (t, x^a)$ and $n$ is the unit normal vector to a constant $t$ hypersurface, so $n = \frac{1}{\sqrt{N}} \frac{\partial}{\partial t}$. 
In the ultrastatic frame we have $n' = \frac{\partial}{\partial t}$, $\sqrt{\bar{g}'} = \frac{1}{N} \sqrt{\bar{g}} $ and $T'^{CFT}_{tt} = \sqrt{N} T^{CFT}_{tt}$, so that
\begin{eqnarray}
E' = \int_{\Sigma} \sqrt{\bar{g}'} T'^{CFT}_{AB} n'^A v^B = E \; .
\end{eqnarray}
Thus energy with respect to $\partial / \partial t$ is invariant under the conformal transformation. Since this is the key quantity of interest we see we may work in the ultrastatic frame without loss of generality. Hence from now on we shall only consider the ultrastatic case with $N = 1$, and so the energy $E[ \Sigma ]$ is a functional of the 2-dimensional closed spatial geometry $\Sigma$. 

As we are interested in the zero temperature vacuum state and thermal states, and the CFT spacetime is static, we assume the dual bulk spacetime is also static. Note that while the Killing vector $\partial/\partial t$ of the CFT metric is globally timelike, we do not assume this for the bulk. Instead we make the following assumption. 
\\

\noindent
\fbox{
\parbox{\columnwidth}{
\emph{Assumption 1:} 
At finite CFT temperature, $T$, a thermal state is described by a dual static bulk solution that is smooth away from the conformal boundary (with boundary metric $g_{CFT}$), and ends only on this or on 
smooth Killing horizons (whose Hawking temperatures are $T$ with respect to $\partial / \partial t$).
}
}
\\

We are assuming all horizons are Killing horizons with respect to $\partial / \partial t$ and in equilibrium with each other. Their surface gravity, $\kappa$, measured with respect to the bulk Killing field $K = \partial / \partial t$ that generates these Killing horizons is given by $\kappa^2 = - \frac{1}{2} \left( \nabla^{(4)}_\mu K_\nu \right)^2$ evaluated at the horizon. With respect to $K$ they have Hawking temperature $T_{Hawking} = \kappa /(2 \pi)$, and since we have chosen a conformal frame for $g_{CFT}$ such that $K$ restricted to the conformal boundary generates the CFT time translation then the CFT temperature is $T = T_{Hawking}$.

The zero temperature vacuum (ie. Casimir) energy is the quantity we are ultimately interested in. One might expect that given the above assumption, we would further assume that if the bulk spacetime at zero temperature has any other ends than the conformal boundary, these should be smooth extremal (ie. zero temperature) horizons. However this would be too restrictive and rules out consideration of important physical examples. 

Let us now consider such an example, namely the dual to the CFT on a flat spatial torus. Taking the CFT to be at temperature $T$ on a boundary metric which is the Minkowski spacetime, the dual bulk is of the form,
\begin{eqnarray}
ds^2_{(4)} = \frac{\ell^2}{z^2} \left( - f dt^2 + \delta_{ab} dx^a dx^b + \frac{1}{f} dz^2 \right) .
\end{eqnarray}
At zero temperature this is Poincare-AdS with $f = 1$, and ends in an extremal horizon as $|x^a|, z \to \infty$ \cite{Kunduri:2013gce}.
At finite temperature it is planar AdS-Schwarzschild with $f = 1 - \left(\frac{z}{z_0}\right)^3$ ending on a regular horizon at $z = z_0$, corresponding to CFT temperature $T = 3 / ( 4 \pi  z_0 )$. 
However the spatial section of the Minkowski spacetime is $\mathbb{R}^2$ and is of course not closed. Since we are interested in closed $\Sigma$ we may rectify this by the identification of Minkowski to the product of time with a 2-torus, ie. making the $x^a$ coordinates above periodic, so $g_{CFT} = -dt^2 + g_{T^2}$ for $g_{T^2}$ the metric on the torus. While this identification on the planar AdS-Schwarzschild solution simply compactifies the geometry of the horizon at $z = z_0$ to be toroidal, at zero temperature it has a more dramatic action, destroying the smooth extremal horizon and replacing it with a null singularity. 
Note that if we consider periodic boundary conditions for fermions about the $x^a$ circles then this is the relevant dual to describe the vacuum (the non-singular AdS-soliton is not a possible bulk geometry since it is incompatible with the fermion spin structure \cite{Horowitz:1998ha}).
This bulk null singularity is a `good' singularity in the sense of \cite{Gubser:2000nd} since at any small finite temperature compared to the torus periods it is shrouded behind the smooth planar AdS-Schwarzschild horizon. Nonetheless we see that this canonical example of the CFT on the product of time with a 2-torus does not have a smooth bulk at zero temperature. Indeed recently it has been argued that extremal horizons are actually non-generic as ends to zero temperature bulk duals, and in fact null singularities might generally be expected \cite{Hickling:2014dra,Hickling:2015ooa}.

Fortunately the result we will derive concerning the zero temperature behaviour does not require such a strong assumption as the bulk ending only on the conformal boundary or smooth extremal horizons. In fact our result requires only that the zero temperature bulk solution arises as the limit of finite temperature solutions of our assumed form above as $T \to 0$. The condition on these is not that the bulk geometry is smooth in this limit, but rather that the total energy and entropy are finite and continuous in this limit. Thus we make the following assumption.
\\

\noindent
\fbox{
\parbox{\columnwidth}{
\emph{Assumption 2:} At zero temperature the static bulk solution dual to the CFT vacuum is the limit of finite temperature solutions satisfying Assumption 1 above. The energy and entropy are well behaved in the $T \to 0$ limit.
}
}
\\

Thus we allow singularities in the gravity dual to the zero temperature vacuum state, but only ones that are {`good'} in the sense of \cite{Gubser:2000nd}.
As emphasised above this allows us to consider situations such as the canonical example of $g_{CFT} = -dt^2 + g_{T^2}$ (with periodic fermion boundary conditions) where the dual vacuum geometry is singular.
Physically if Assumption 2 did not hold one would be concerned the CFT was pathological - indeed even having non-zero entropy as $T \to 0$ is rather non-generic from the perspective of quantum field theory. 
Thus we will always implicitly be working at finite temperature in what follows, and will talk about the zero temperature behaviour only as a limit of this.

Following our previous work \cite{Hickling:2015sha} we write the static bulk spacetime in terms of a warped product of time and a Riemannian geometry $(\mathcal{M},g)$,
\begin{eqnarray}
\label{eq:bulkmetric}
ds^2_{(4)} = \frac{\ell^2}{Z^2} \left( -dt^2 + g_{ij} dx^i dx^j \right)
\end{eqnarray}
where $x^i$ are local coordinates on $\mathcal{M}$, so $i=1,2,3$. This geometry $(\mathcal{M},g)$ is known as the \emph{optical geometry} \cite{OptMetric}.
We assume that the bulk spacetime ends on the conformal boundary, with geometry $(\mathbb{R} \times \Sigma, g_{CFT})$, with $g_{CFT} = - dt^2 + \bar{g}$, and may also end on $N_H \ge 0$  bulk Killing horizon components, with spatial sections $\mathcal{H}_A$ for $A = 1, \ldots N_H$.

We take $\mathcal{M}$ to have a boundary $\partial \mathcal{M} = \Sigma$, and make $Z$ the defining function for the conformal boundary, so $Z > 0$ everywhere except on the boundary  $\partial M$, where $Z = 0$ with $d Z \ne 0$. 

Consider a component $\mathcal{H}_A$ of the horizon.
For a general static Killing horizon we may write the metric locally as,
\begin{eqnarray}
ds^2 = - \kappa^2 r^2 f(r,y) dt^2 + dr^2 + h_{ab}(r, y) dy^a dy^b 
\end{eqnarray}
where the constant $\kappa$ gives the surface gravity of the horizon with respect to the Killing vector $\partial/\partial t$, $r$ is a normal coordinate to the horizon which is located at $r = 0$, $y^a$ are coordinates on the spatial section of the horizon, and $f$ is a function such that $f(0,y) = 1$, and both $f(r,y)$ and $h_{ab}(r,y)$ are smooth functions of $r^2$ to ensure regularity of the Killing horizon  (see for example \cite{Wiseman:2011by,Adam:2011dn}). 
For the dual to a thermal state, the surface gravity of such a bulk horizon is related to the CFT temperature as $T = \kappa/2 \pi$, and the CFT entropy contribution due to this horizon component is $S_{\mathcal{H}_A} = A_{\mathcal{H}_A}/4 G_{(4)}$ where $A_{\mathcal{H}_A} = \int \sqrt{h}|_{r=0}$ is the area of the horizon $\mathcal{H}_A$. The AdS/CFT dictionary then states that the total CFT entropy, $S$, is the sum of the horizon components so $S = \sum_A S_{\mathcal{H}_A}$.

Written in terms of our optical metric we see that near a horizon we have, 
\begin{align}
\label{eq:horizon1}
g_{ij} dx^i dx^j &= \frac{1}{\kappa^2 r^2 f} \left( dr^2 + h_{ab}(r, y) dy^a dy^b \right) \nonumber \\
Z &= \frac{\ell}{\kappa r \sqrt{f}} \; .
\end{align}
Hence from the perspective of the optical geometry $(\mathcal{M}, g)$ we see as $r \to 0$ each horizon component is actually an asymptotic region that is in fact a conformal boundary, with conformal boundary geometry given by the geometry of the spatial section of $\mathcal{H}_A$. Of course these are conformal boundaries of $(\mathcal{M},g)$ and are not to be confused with the conformal boundary of the full spacetime which, as we have discussed, corresponds to an actual boundary of $\mathcal{M}$.

In summary the assumed structure of $(\mathcal{M}, g)$ for bulk spacetimes at finite temperature is as follows. The spacetime conformal boundary where $Z = 0$ is a \emph{boundary} of $\mathcal{M}$ with $\partial \mathcal{M} = \Sigma$. The $N_H$ bulk spacetime Killing horizons where $Z \to \infty$ are \emph{conformal boundaries} of $\mathcal{M}$ with geometries given by the spatial sections of the horizons $\mathcal{H}_A$.

%-----------------------------------------------------------------
%
\section{Bulk equations
\label{sec:bulk}}
%
%-----------------------------------------------------------------

The static Einstein equations for this bulk spacetime can be decomposed over the optical geometry as,
\begin{eqnarray}
\label{eq:bulkeqns}
R_{ij} &=& - \frac{2}{Z} \nabla_i \partial_j Z \nonumber \\
  R &=& \frac{6}{Z^2} \left( 1 - \left( \partial Z \right)^2 \right)
\end{eqnarray}
where indices are raised/lowered using the optical metric $g_{ij}$. Here $R_{ij}$ is the Ricci tensor of the optical metric $g_{ij}$, $R = R_i^{~i}$ is its Ricci scalar and $\nabla$ is its covariant derivative. 

Our results all follow from the elegant elliptic equation that the optical Ricci scalar obeys,
\begin{eqnarray}
\label{eq:optRicciEq}
\nabla^2 R -  R^2 + 3 R_{ij} R^{ij} = 0
\end{eqnarray}
which may be verified straightforwardly from the Einstein equations above.
Since the norm in $(\mathcal{M},g)$ of the tracefree part of the optical Ricci tensor $\tilde{R}_{ij} \equiv {R}_{ij} - \frac{1}{3} R g_{ij}$ is given as,
\begin{eqnarray}
 | \tilde{R}_{ij}  |^2 = R_{ij} R^{ij} - \frac{1}{3} R^2 
\end{eqnarray}
and is non-negative for  a smooth Riemannian optical metric,
we see that $R_{ij} R^{ij} \ge \frac{1}{3} R^2$. 
Hence the optical Ricci scalar is a super-harmonic function on $(\mathcal{M}, g)$,
\begin{eqnarray}
\nabla^2 R \le 0 
\end{eqnarray}
with this inequality being saturated only if $\tilde{R}_{ij}$ vanishes.
This result played a key role in our previous work \cite{Hickling:2015sha}.
Here we integrate it over the optical geometry, and use the divergence theorem to obtain,
\begin{eqnarray}
\label{eq:surfterms}
\int_{\mathcal{M}} \sqrt{g} \, \nabla^2 R = \int_{\partial \mathcal{M}} dA^i \partial_i R + \sum_A \int_{\mathcal{H}_A} dA^i \partial_i  R \le 0 \qquad
\end{eqnarray}
where $dA^i$ is the outward facing area element for a surface. Thus we find an inequality involving surface terms over the boundary of $\mathcal{M}$, 
corresponding to the spacetime conformal boundary, and also the asymptotic regions of $\mathcal{M}$ corresponding to the spacetime horizons. Having this inequality we must now evaluate these surface terms.

%-----------------------------------------------------------------
%
\section{Conformal boundary surface term
\label{sec:boundary}}
%
%-----------------------------------------------------------------

Firstly we consider the surface term at $\partial \mathcal{M}$ due to the conformal boundary and relate this to physical quantities using the holographic dictionary \cite{Henningson:1998gx,Balasubramanian:1999re,deHaro:2000xn} (and reviewed in \cite{Skenderis:2002wp}).
The Fefferman-Graham form for a conformally compact $(3+1)$-dimensional Einstein metric is,
\begin{eqnarray}
ds^2_{(4)} = \frac{\ell^2}{z^2} \left(  dz^2 + h_{AB}(z, x) dx^A dx^B \right)
\end{eqnarray}
with the asymptotic behaviour near the conformal boundary $z = 0$ given by,
\begin{align}
h_{AB}(z, x) & = \bar{h}_{AB}(x) + \left( \bar{R}^{(h)}_{AB} - \frac{1}{4} \bar{R}^{(h)} \bar{h}_{AB} \right) z^2 \nonumber \\
& \qquad + t_{AB}(x) z^3 + O(z^4)
\end{align}
where this series expansion is written with indices raised and lowered with respect to the metric that provides the representative for the conformal boundary geometry, $\bar{h}_{AB}(x)$. In particular $\bar{R}^{(h)}_{AB}$ is its Ricci tensor, and $\bar{R}^{(h)}$ its scalar curvature. The data $\bar{h}_{AB}(x)$ and $t_{AB}(x)$, which is a transverse traceless tensor with respect to $\bar{h}_{AB}$, fully determine all subsequent terms in the above expansion. 
The AdS/CFT dictionary then requires us to take $\bar{h} = g^{CFT} = -dt^2 + \bar{g}$, and then the vacuum expectation value of the CFT stress tensor, $T_{CFT}$, is,
\begin{eqnarray}
\langle T^{CFT}_{AB} \rangle = 3 \, c \, t_{AB}
\end{eqnarray}
where $c$ is the effective central charge defined above.
Hence given a bulk solution with such asymptotics, the CFT energy is then,
\begin{eqnarray}
E = \int_{\Sigma} \sqrt{\bar{g}} \langle T^{CFT}_{tt} \rangle  =  3 \, c \, \int_{\Sigma} \sqrt{\bar{g}} \, t_{tt} \; .
\end{eqnarray}
 It is worth emphasising that due to the absence of a conformal anomaly there is no ambiguity or scheme dependence in this stress tensor. 
 
Consider our metric in equation~\eqref{eq:bulkmetric}. Taking coordinates on $\mathcal{M}$ so $x^i = ( z, x^a)$ with $z$ the Feffermann-Graham coordinate above, then we see near the conformal boundary,
\begin{align}
Z(z,x) & = z  \left( 1 - \frac{1}{8} \bar{R}(x) z^2 + \frac{1}{2} t_{tt}(x) z^3 + O(z^4) \right) \nonumber \\
g_{zz}(z,x) & = 1 - \frac{1}{4} \bar{R}(x) z^2 + t_{tt}(x) z^3+ O(z^4)
\nonumber \\
g_{ab}(z,x) & = \bar{g}_{ab}(x) - \frac{1}{2} \bar{R}(x) \bar{g}_{ab}(x) z^2 \nonumber \\
& \qquad + \left( t_{ab}(x) + \bar{g}_{ab}(x) t_{tt}(x) \right) z^3 + O(z^4)
\end{align}
with $g_{z a} = 0$, and
where now the expansions are written covariantly with respect to the CFT spatial metric $\bar{g}$, with its Ricci tensor and scalar being $\bar{R}_{ab}$ and $\bar{R}$ respectively. Using this we may compute the asymptotic behaviour of the optical Ricci scalar,
\begin{align}
R(z, x) & = 3 \bar{R}(x) - 18 t_{tt}(x) z + O( z^2) \; .
\end{align}
Note  that this implies  the Ricci scalar of $\Sigma$ is simply related to the boundary value of the bulk optical Ricci scalar as \cite{Hickling:2015sha},
\begin{align}
\label{eq:boundaryrelation}
\bar{R} = \frac{1}{3} R |_{\partial M} \; .
\end{align} 
Now we may compute the boundary term $\int_{\partial \mathcal{M}} dA^i \partial_i  R$. Using the above expansions we have,
\begin{eqnarray}
\partial_n R & =  18 t_{tt} + O( z)
\end{eqnarray} 
where $n = - \frac{1}{\sqrt{g_{zz}}} \frac{\partial}{\partial z}$ gives the unit normal to a constant $z$ surface in $(\mathcal{M},g)$ directed towards the conformal boundary, so that,
\begin{align}
\label{eq:surfasym}
\int_{\partial \mathcal{M}} dA^i \partial_i  R & = \int_{Z = 0} \sqrt{\bar{g}} \, \partial_n R  \nonumber \\
& =  18  \int_{\Sigma}  \sqrt{\bar{g}} \, t_{tt} = \frac{6}{c} E \; .
\end{align}
Thus we see that the surface term associated to the spacetime conformal boundary is simply  proportional to the CFT energy.

We note that if a finite or zero temperature bulk spacetime ends only on the conformal boundary, with no horizons or singularities, then the only boundary term is the one above and from~\eqref{eq:surfterms} this simply yields the result $E \le 0$. Thus we can already see that the energy in these cases is non-positive.

An example of such a situation is for a CFT  where $\Sigma$ is a round sphere radius $\mathcal{R}$, and the  temperature is taken to be well below that of the Hawking-Page phase transition \cite{Hawking:1982dh,Witten:1998zw}. In this case it is expected that no static black hole solutions exist at such temperatures (certainly this is true imposing spherical symmetry) and the only bulk is global AdS. Thus taking $\Sigma$ as a small deformation of a round sphere and taking low temperatures, so $T \mathcal{R} \ll 1$ we expect no bulk horizons and hence again $E \le 0$ by the above. Another example is for $\Sigma$ a 2-torus with antiperiodic fermion boundary conditions about one cycle. The relevant static bulk is then the AdS-soliton  \cite{Horowitz:1998ha}. It is believed that black holes have a minimum temperature \cite{Aharony:2005bm,Emparan:2009dj,Figueras:2014lka} with such asymptotics, and so below this temperature the only candidate bulk spacetime is the AdS-soliton itself which indeed has negative energy $E < 0$. More generally we might imagine that boundary metrics $\Sigma$ that lead to confining behaviour at low temperatures (for example, taking $\Sigma$ to have positive scalar curvature \cite{Hickling:2015sha}) have dual static geometries with no bulk horizons below some threshold temperature, and hence in such a temperature range must have $E \le 0$.

However, as emphasised above, in general we may have bulk horizons at arbitrarily low temperature as in the example of $\Sigma$ being a torus (with periodic fermion boundary conditions). Hence we now proceed to consider the contribution to surface terms from finite temperature horizons in the bulk.

%-----------------------------------------------------------------
%
\section{Horizon surface terms
\label{sec:horizon}}
%
%-----------------------------------------------------------------

A static smooth spacetime Killing horizon can be written in local coordinates as in equation~\eqref{eq:horizon1} in our bulk ansatz. In order to compute the surface term due to a  horizon component $\mathcal{H}_A$ we solve the bulk Einstein condition as an expansion in the normal coordinate $r$ about the horizon location $r=0$. One finds,
\begin{align}
f(r,y) & = 1 - \frac{1}{6} \bar{R} r^2 + O(r^3) \nonumber \\
h_{ab}(r, y) &= \bar{h}_{ab} + \left( \frac{3}{2 \ell^2} + \frac{1}{4} \bar{R} \right) \bar{h}_{ab} r^2 + O(r^3)
\end{align}
where $\bar{h}_{ab}(y)$ is the metric induced on the spatial section of the horizon, and $\bar{R}$ is its Ricci scalar.
From this one may deduce that the optical Ricci scalar behaves as,
\begin{align}
\partial_n R & = - \kappa^3 \left( \frac{12}{\ell^2} + 6 \bar{R} \right) r^2 + O(r^3)
\end{align}
where $n = - \frac{1}{\sqrt{g_{rr}}} \frac{\partial}{\partial r}$  gives a unit normal to a constant $r$ surface in $(\mathcal{M},g)$ directed into the horizon,
and then taking into account the fact that the volume element of a constant $r$ surface scales as $\sim \frac{1}{\kappa^2 r^2} \sqrt{\bar{h}}$ as $r \to 0$,  one finds the finite value,
\begin{align}
\int_{\mathcal{H}_A} dA^i \partial_i  R & = \int_{r = 0} \sqrt{\bar{g}} \partial_n R  \nonumber \\
& = - \kappa  \int \sqrt{\bar{h}} \left( \frac{12}{\ell^2} + 6 \bar{R} \right) \nonumber \\
& = - \frac{12 \kappa}{\ell^2} A_{\mathcal{H}_A} - 6 \kappa  \int \sqrt{\bar{h}} \bar{R} \; .
\end{align}
Now using the relations to CFT temperature $T$, the contribution to the entropy, $S_{\mathcal{H}_A}$, and the Gauss-Bonnet theorem, we may write,
\begin{align}
% typo corrected
\int_{\mathcal{H}_A} dA^i \partial_i  R 
& = - \frac{6 }{c} T S_{\mathcal{H}_A} - 48 \pi^2 \chi_{\mathcal{H}_A} T  
\end{align}
where $\chi_{\mathcal{H}_A}$ is the Euler characteristic for the spatial horizon geometry of $\mathcal{H}_A$.

Since all the horizons are in equilibrium at the same temperature then the total contribution due to the horizons is,
\begin{align}
\label{eq:surfhoriz}
% typo corrected
\sum_A \int_{\mathcal{H}_A} dA^i \partial_i  R 
& = - \frac{6 }{c} T S - 48 \pi^2 T \sum_A \chi_{\mathcal{H}_A}  
\end{align}
with $S = \sum_A S_{\mathcal{H}_A}$ being the CFT entropy.
\footnote{It is a simple matter to allow the horizons to have their own temperatures, and hence not be in equilibrium, but here we are interested in the equilibrium canonical ensemble.}

%-----------------------------------------------------------------
%
\section{Vacuum energy bounds
\label{sec:result}}
%
%-----------------------------------------------------------------

Thus we see that the inequality $\nabla^2 R \le 0$ in the optical Ricci scalar integrated over the bulk yields the bound on the surface terms in~\eqref{eq:surfterms} which may be evaluated using equations~\eqref{eq:surfasym} and~\eqref{eq:surfhoriz} to deduce,
\begin{align}
\label{eq:thermobound}
% typo corrected
\frac{1}{c} F = \frac{1}{c} \left( E -  T S  \right)\le 8 \pi^2 \, T \sum_A \chi_{\mathcal{H}_A}  
\end{align}
where $F = E - T S$ is the CFT free energy at temperature $T$.
While this bound can be thought of as a constraint on the thermodynamics of the CFT and equivalently on the dual black holes, it also allows us to bound the zero temperature vacuum Casimir energy by taking the limit $T \to 0$. By Assumption 2 the total energy and entropy are bounded and continuous in the limit $T \to 0$, so,
\begin{eqnarray}
\lim_{T \to 0} \left( \frac{ E }{c } \right) \le 0
\end{eqnarray}
and hence the vacuum energy is non-positive. 

As discussed above, these inequalities can only be saturated if the tensor $\tilde{R}_{ij}$ (the tracefree part of $R_{ij}$) vanishes everywhere in $\mathcal{M}$. However, since,
\begin{eqnarray}
\nabla^i \tilde{R}_{ij} = \nabla^i {R}_{ij} - \frac{1}{3} \partial_j R =  \frac{1}{6} \partial_j R
\end{eqnarray}
 we see that a necessary condition for saturation of these bounds is that the optical Ricci scalar is constant on $\mathcal{M}$. Furthermore equation~\eqref{eq:boundaryrelation} then implies that $\bar{R}$, the Ricci scalar of $\Sigma$, must also be constant.
We conclude  that static bulk spacetimes where $\Sigma$ has non-constant scalar curvature must have negative vacuum energies $E < 0$. Thus the vacuum energy can only vanish in the non-generic situation that $\Sigma$ has constant curvature, and hence is locally a sphere, torus or hyperbolic space.

%-----------------------------------------------------------------
%
\section{Discussion
\label{sec:discussion}}
%
%-----------------------------------------------------------------

We now conclude with a brief discussion. Firstly we point out that since our result only applies in the large $c \to \infty$ limit (required for having a gravity dual), it is only the leading part of the energy $E$ that is constrained to be non-positive. However holography predicts that provided a bulk dual exists, the vacuum energy will generically go as $E \sim O(c)$ (except, as we have seen, in cases of maximally symmetric $\Sigma$ where it may vanish to leading order in $c$). We emphasise that our results above rely on the CFT having a dual description given by a bulk spacetime which is smooth at any small finite temperature. If it happened that for some $\Sigma$ no bulk spacetime existed, or somehow violated our assumptions, then the result is not expected to hold, although we know of no such situation.

We note that the thermodynamic bound~\eqref{eq:thermobound} is precisely the one found previously in \cite{Galloway:2015ora} specialised to the case of a $(3+1)$-dimensional bulk although in that work $\Sigma$ was constrained to be of constant curvature, so a round sphere, flat torus (or quotients of these) or compact hyperbolic space.\footnote{
Note that in these cases of $\Sigma$ we already know explicit bulk metrics both at finite and zero temperature (although we certainly do not know all such solutions).
}
It is interesting to note that the bound in \cite{Galloway:2015ora} was given in any bulk dimension.
For $D$ boundary dimensions, and hence $(D+1)$ bulk dimensions, the optical Ricci scalar obeys the relation \cite{Hickling:2015sha}, 
\begin{eqnarray}
\nabla^i \left( \frac{1}{Z^{D-3}} \partial_i R \right) \le 0 \; .
\end{eqnarray}
However, in the higher dimensional cases $D > 3$ this inequality does not obviously lead to analogous bounds on the thermodynamics or energy. Integrating and using the Gauss law over $(\mathcal{M}, g)$ the surface term generated due to the bulk spacetime conformal boundary is no longer generally finite. It would obviously be interesting to explore whether our arguments can be modified to yield bounds in higher dimensions too. Another interesting question is whether these results generalise to inclusion of bulk matter fields obeying specific energy conditions. 

It is worth contrasting our result with previous work on positive energy theorems \cite{Gibbons:1983aq,Cheng:2005wk} relevant for the AdS-CFT setting. In the $(2+1)$-dimensional case we consider it was argued in \cite{Cheng:2005wk} that for time dependent bulk dynamics a necessary condition for the energy to be bounded from below by zero was the existence of a spinor on $\Sigma$ obeying certain differential conditions. 
On the other hand our result shows that for generic $\Sigma$ we expect the energy to be negative for the bulk vacuum geometry, and it must therefore be true that the assumptions required for stability as discussed in \cite{Cheng:2005wk} cannot hold. 
It is important to note that this does not imply such a setting is necessarily dynamically unstable.
Indeed the example of the AdS-soliton \cite{Horowitz:1998ha} is believed to be stable to bulk perturbations.

We conclude by emphasising that $(2+1)$-dimensional holographic CFTs are known explicitly, as in the canonical example of the ABJM theory \cite{Aharony:2008ug}.
It is interesting to note that they may potentially be of interest in the context of AdS/CMT \cite{Hartnoll:2009sz} where one might in principle imagine a real world experiment simulating a holographic $(2+1)$-dimensional CFT, where the spatial geometry of the material could be deformable away from a very symmetric case such as a plane, leading to a Casimir effect. 
Since our results show the energy is reduced to become negative for generic perturbations of flat space, this indicates potential instabilities  associated to crumpling driven by a decrease in the vacuum energy.

%-----------------------------------------------------------------
%
\section{Relation to Anderson's bound
\label{sec:Anderson}}
%
%-----------------------------------------------------------------

After completion of this work Juan Maldacena pointed out to us another way that the above results may be derived using the geometric bound of Anderson \cite{Anderson}. We have included a sketch of this alternate derivation for the interest of the reader noting that it provides a beautiful example of how a physical result may be geometrized in AdS-CFT.

Starting from the 4-dimensional Gauss-Bonnet theorem Anderson has shown that the renormalized volume of an asymptotically hyperbolic Einstein manifold satisfies the bound,
\begin{align}
V_{ren} \le \frac{4 \pi^2}{3} \ell^4 \chi(M_4) \; .
\end{align}
Here $\chi(M_4)$ is the Euler character of the bulk spacetime, calculated by writing this as a conformally compact manifold, taking $ds^2_4 = \frac{1}{Z^2} g_{AB} dx^A dx^B$ for a defining function $Z$ and taking $(M_4, g)$ to be a smooth Riemannian manifold with boundary $\partial M_4$ where the defining function vanishes.
This bound is saturated if and only if the bulk Weyl tensor vanishes.

Since we are considering static bulk spacetimes with horizons at equilibrium, we may analytically continue the Lorentzian bulk geometry to a Riemannian metric by continuing to Euclidean time $\tau = i t$. Then with the appropriate identification $\tau \sim \tau + \beta$ the Lorentzian horizons continue to smooth fixed point sets under the Killing symmetry generated by $\partial/\partial \tau$ referred to as `bolts'. 

Following \cite{Maldacena:2011mk} then the renormalized volume is simply related to the renormalized on-shell Euclidean gravitational action, $S_{E,\mathrm{on-shell}} = \frac{3}{8 \pi G \ell^2} V_{ren}$ \cite{Balasubramanian:1999re,Skenderis:2002wp}. Now Euclidean semi-classical gravity implies the Euclidean action (suitably renormalized) is related to the free energy $F$ as, $S_{E,\mathrm{on-shell}} = \frac{1}{T} F$. Via Anderson's bound this yields,
\begin{align}
\label{eq:newbound}
\frac{1}{T} F \le 8 \pi^2 c \, \chi(M_4)
\end{align}
This is similar in spirit to our thermodynamic bound in equation~\eqref{eq:thermobound} although the Euler number refers to the bulk spacetime rather than horizons that might be present. Note that following our argument and assumptions in the paper then considering this bound at finite temperature and then taking $T \to 0$, so $F \to E$,  this implies $E \le 0$ (again without assuming the zero temperature solution itself is smooth). 

However we may also recover our thermodynamic inequality~\eqref{eq:thermobound} directly from the bound above. Following \cite{Gibbons:1979xm} then the Euler character of a 4-manifold with a Killing vector may be given in terms of the nature of the fixed point sets of the action generated by this, in particular the number of `nuts', `anti-nuts' and `bolts'. This holds for manifolds with boundary provided that the Killing vector is tangential to the boundary. In our case the defining function has been chosen to be compatible with the static symmetry so that $(M_4,g)$ is a static smooth 4-manifold with boundary, and Killing vector $\partial/\partial \tau$ that is tangential to the boundary $\partial M_4$. Then we have,
\begin{align}
\chi(M_4) = N_+ + N_- + \sum_{A=1}^{N_B} \chi_A
\end{align}
where $N_{+}$ and $N_-$ are the number of nuts and anti-nuts, and $\chi_A$ is the Euler character of the $A$-th bolt, of which there are $N_B$. In our application we have no nuts or anti-nuts, and the bolts continue to the horizons, with the 2-geometry of the bolt corresponding to the spatial section of the Lorentzian horizon. Hence for our application this implies,
\begin{align}
\chi(M_4) = \sum_{A} \chi({\mathcal{H}_A})
\end{align}
and substituting into~\eqref{eq:newbound} yields our bound~\eqref{eq:thermobound}.

Anderson's bound is saturated for vanishing bulk Weyl tensor whereas we had previously found saturation for vanishing $ \tilde{R}_{ij} $, the traceless Ricci tensor of the optical metric. A simple computation confirms that the 4-dimensional 

%-----------------------------------------------------------------
%
\section*{Acknowledgements}
%
%-----------------------------------------------------------------

We thank Philip Candelas and Paul McFadden for useful comments. 
We are very grateful to Juan Maldacena for pointing out the relation to Anderson's bound which we have detailed in section \ref{sec:Anderson}.
This work was supported by the STFC grant ST/J0003533/1. AH is supported by an STFC studentship.

%-----------------------------------------------------------------
%
\bibliographystyle{apsrev4-1}
\bibliography{paperCasimirV1}
%
%-----------------------------------------------------------------

\end{document}